\documentclass[12pt]{iopart}

\newcommand{\eq}[1]{(\ref{#1})}
\usepackage{graphicx}

\begin{document}

\title[Kerr Naked Singularities as Particle Accelerators]{Kerr Naked Singularities as Particle Accelerators}

\author{Mandar Patil and Pankaj Joshi}

\address{Tata Institute of Fundamental Research, Homi Bhabha Road,
Mumbai 400005, India}
\ead{mandarp@tifr.res.in}
\begin{abstract}
We investigate here the particle
acceleration by Kerr naked singularities. We consider
a collision between particles dropped in from infinity
at rest, which follow geodesic motion in the equatorial plane,
with their angular momenta in an appropriate finite range
of values. When an event horizon is absent, an initially
infalling particle turns back as an outgoing particle,
when it has the angular momentum in an appropriate range
of values, which then collides with infalling particles. When
the collision takes place close to what would have been
the event horizon in the extremal case, the center of mass
energy of collision is arbitrarily large, depending on
how close is the overspinning Kerr geometry to the
extremal case. Thus the fast rotating Kerr configurations
if they exist in nature could provide an excellent
cosmic laboratory to probe ultra-high-energy physics.

\end{abstract}

\maketitle

\section{Introduction}

An intriguing possibility to study new physics at
ultrahigh energies, which remains unexplored by terrestrial
particle accelerators, is to make use of naturally occurring
astrophysical exotic objects. In this spirit, the divergence
of center of mass energy of infalling particles
colliding near the event horizon of near extremal Kerr
blackholes, and the observational signatures of such a
process in the context of dark matter annihilations
were studied recently
\cite{BSW} \cite{BSW2}.
However, this process
suffers from several drawbacks, such as the extreme
fine-tuning required of the angular momentum of the particles,
and also an
infinite proper time needed for the collision
events to take place
\cite{Berti}.

In this note we investigate the particle
collision with ultrahigh energies in the background of
near extremal Kerr naked singularities, transcending
the Kerr bound by vanishingly small amount.
The interesting point that we show here is, in such a
process the drawbacks mentioned
above are naturally circumvented, due to the absence of
an event horizon. This allows us to consider high energy
collisions between ingoing and outgoing particles in
a generic manner, unlike in the blackhole case.

\section{Geodesics in Kerr geometry in equatorial plane}

We examine here the particle
collisions in the background of a Kerr naked singularity.
For simplicity and clarity we focus
on the test particles following timelike geodesics
in the equatorial plane. The Kerr metric
\cite{Kerr}
in Boyer-Lindquist
coordinates $\left(t,r,\theta,\phi\right)$
in the equatorial plane, $\theta=\frac{\pi}{2}$ is given by
\begin{equation}
ds^2=- \left( 1- \frac{2}{r}\right)dt^2- \frac{4a}{r}
dt d\phi
+\left(\frac{r^2}{\Delta}\right)dr^2+ r^2 d\theta^2+
\left(r^2+a^2+\frac{2a^2}{r}\right) d\phi^2 \label{KBL1}
\end{equation}
where $\Delta= r^2+a^2-2r$. We work in the units
$c=G=M=1$, $M$ being the mass and $a$ the angular momentum
parameter. The event horizon is obtained by solving
$\Delta=0$. It then follows that when $a>1$, there is no event
horizon and the timelike naked singularity at
$r=0,\theta=\frac{\pi}{2}$ is exposed to asymptotic
observers. We also note that for extremal Kerr
blackhole, namely with $a=M=1$, the event
horizon is located at $r=1$.

The metric admits killing vectors
$\partial_{t},\partial_{\phi}$ and thus the quantities
\begin{eqnarray}
\nonumber
E=-g_{\mu\nu}\left(\partial_{t}\right)^{\mu} U^{\nu}
=-g_{tt}U^{t}-g_{t\phi}U^{\phi} \\
\nonumber
L= g_{\mu\nu}\left(\partial_{\phi}\right)^{\mu} U^{\nu}
=-g_{\phi}U^{t}-g_{\phi\phi}U^{\phi}
\end{eqnarray}
are conserved along the
geodesics. These are interpreted as conserved energy
and angular momentum per unit mass of the particle, $U$
being the four
velocity of the particle. Solving for
$U^t$,$U^{\phi}$ we get,
\cite{BSW},\cite{Bardeen},
\begin{equation}
U^{t}=\frac{1}{\Delta}\left[\left(r^2+a^2+\frac{2 a^2}{r}\right)E
-\frac{2a}{r} L \right]
=\frac{1}{r^2}\left[-a\left(aE-L\right)+\frac{r^2+a^2}{\Delta}T\right] \label{Ut}
\end{equation}
\begin{equation}
U^{\phi}= \frac{1}{\Delta} \left[\left(1-\frac{2}{r}\right)L
+\frac{2a}{r}E\right]
=\frac{1}{r^2}\left[\left(L-aE\right)+\frac{a}{\Delta}T\right] \label{Up}
\end{equation}
where $T=E\left(r^2+a^2\right)-La$.

From \eq
{Ut},\eq{Up}, $U^{\theta}=0$ and normalization
condition $U^{\mu}U_{\mu}=-1$ for a timelike geodesic,
the radial component of velocity can be written as,
\begin{eqnarray}
\nonumber
U^{r}=\pm\sqrt{E^2-1+\frac{2}{r}-\frac{\left(L^2-a^2
\left(E^2-1\right)\right)}{r^2}+
\frac{2\left(L-aE\right)^2}{r^3}} \\
=\pm \frac{1}{r^2}\sqrt{T^2-\Delta\left(r^2+\left(L-aE\right)^2\right)} \label{Ur}
\end{eqnarray}
Here $\pm$ stand for radially outgoing and
ingoing geodesics respectively.
The above equation can be cast in the form
\begin{eqnarray}
\nonumber
U^{r 2 }+V_{eff}(L,E,r)=0 \\
V_{eff}=-E^2+1-\frac{2}{r}+\frac{\left(L^2-a^2
\left(E^2-1\right)\right)}{r^2}-\frac{2\left(L-aE\right)^2}{r^3}
\label{Veff}
\end{eqnarray}
where $V_{eff}(L,E,r)$ can be thought
of as an effective potential for the radial motion.
The center of mass energy of collision
\cite{BSW}
of two particles with velocities $U_{1}$ and $U_{2}$
is given by
\begin{equation}
 E_{c.m.}^2=2m^2\left(1-g_{\mu\nu}U_{1}^{\mu}U_{2}^{\nu}\right)
\label{Ecm}
\end{equation}
We restrict our attention here to the geodesics with
conserved energy per unit mass $E=1$, corresponding
to the case of marginally bound particles, released
at infinity from rest, whose energy comes solely from
the gravitational acceleration of the Kerr spacetime.

\section{Particle acceleration for extremal Kerr blackholes}

Towards considering the particle collisions in
the Kerr geometry, we first note that in the Banados-Silk-West
(BSW) mechanism of particle acceleration by near-extremal
Kerr blackholes, two identical particles at rest
each with mass $m$ are released from infinity, and
are made to collide near the horizon of the Kerr black hole.
These ingoing particles are highly blue-shifted
by the time they reach the event horizon. But in most of
the cases they reach the horizon almost perpendicular to it,
so the relative velocity of approach of two particles
happens to be small. Therefore the center of mass energy of particles
is finite and not significantly larger than their
rest mass energy. Thus particles which participate in collisions
must have large and opposite angular momenta, so as to
maximize the relative velocity of collision between the particles.
Particles with small angular momentum fall into the blackhole
almost perpendicular to the horizon, whereas particles with
rather large angular momenta turn back even before they could reach
the horizon. It follows that one must then fine-tune the
angular momentum of a particle to a largest possible value
that still makes it possible for it to reach the event
horizon. The center of mass energy of collision is
then maximized in that case.

When the black hole is close to extremality,
this fine-tuned angular momentum approaches a value
$L=\Omega_{H}^{-1}$, where $\Omega_{H}$ is the angular
velocity of the event horizon, and the following
condition is also satisfied,
\begin{equation}
V_{eff}=\frac{dV_{eff}}{dr}=0
\label{Veffz}
\end{equation}
This essentially implies that the particle
travels almost parallel to the event horizon, which is
a null surface, and thus it is ultra-relativistic with
respect to the other particle with which it collides.
This leads to the divergence of the center of mass energy
in the BSW mechanism. Here the equation \eq{Veffz} implies
that the proper time required for the particle to reach
the horizon and also for such a collision to take
place approaches infinity.

\section{Particle acceleration by Kerr naked singularities}

The origin of such issues in the case of blackholes
is that the event horizon is a one way membrane. The chosen
location for collisions with divergent center of mass energy
has to be arbitrarily close to the event horizon, because it
is an infinite blueshift surface for the particles approaching it.
When two infinitely blueshifted particles
collide near the event horizon of the blackhole
with sufficiently large relative velocities, the center
of mass energy of collision is bound to diverge.
In in case of a Kerr black hole, it is not
possible to have a collision between the ingoing and outgoing
particles due to the absence of the outgoing particles near the
event horizon which is a one way membrane in the spacetime.
Therefore one must consider collision between only the infalling
particles near the horizon towards the purpose of
high energy collisions. In such a case, the only way
to maximize relative velocity between them is to
fine-tune the angular momentum of one of the particles.

Such a problem is naturally circumvented
if we consider a near-extremal Kerr naked singularity,
rather than a near-extremal Kerr blackhole.
In such a case there is no event horizon
existing in the spacetime,
thus allowing for the possibility of collision between
an {\it ingoing} and another {\it outgoing} particles,
as shown in Fig1.
Then the relative velocity of collision between these
two particles can be very large, and the requirement
of fine-tuning of angular momentum and various issues
arising from it disappear. As we show later in this
section, the range of allowed particle angular momenta
is a finite interval, unlike the single fine-tuned
value in the blackhole case.

Since the naked singularity is assumed to be near
extremal with
\begin{equation}
 a-M=a-1=\epsilon \rightarrow 0,
\end{equation}
the surface $r=1$, which would have been
the event horizon for the extremal blackhole, is
still a surface with arbitrarily large blueshift
for the particles approaching it.
This follows from the fact that
\begin{equation}
\Delta(r=1) \approx  2\epsilon \rightarrow 0
\label{del}
\end{equation}
Therefore, the center of mass energy of
collision between the two particles, which approach
each other from the opposite directions with large relative
velocity and that suffer from extremely large
blueshift as they approach $r=1$, can be arbitrarily
large.

Let us now consider a collision between two
identical particles of mass $m$, which follow a geodesic
motion along the equatorial plane. The particles are
assumed to be at rest at infinity, so the conserved
energy of each particle is $E=1$.
The effective potential \eq{Veff} for the
radial motion in that case is given by,
\begin{equation}
V_{eff}=-\frac{2}{r}+\frac{L^2}{r^2}-\frac{2\left(L-a\right)^2}{r^3}
\end{equation}
For a particle with the orbital angular momentum
$L=0$, the above expression for the effective potential
implies that the gravity is always attractive in
the equatorial plane (see Fig2). This is unlike the case
where the gravity is repulsive in Kerr geometry, off the
equatorial plane, in the vicinity of naked singularity
\cite{Felice}.
Thus such a particle will fall in with an
ever-increasing radial component of velocity and
eventually hit the naked singularity at $r=0,\theta=\frac{\pi}{2}$.
It then follows that if the ingoing particle were to
turn back, it must have necessarily a non-zero
angular momentum.

An initially ingoing particle will turn back if its effective
potential for radial motion admits a zero.
The radial coordinate where the particle undergoes
a reflection is the larger root of the equation
\begin{equation}
V_{eff}(r)=0
\label{vz}
\end{equation}
which is given by,
\begin{eqnarray}
r=r_{refl}=\frac{L^2}{4}\left[1+\sqrt{D}\right]
\label{rrefl}
\end{eqnarray}
where $D=1-16\frac{\left(L-a\right)^2}{L^4}$.
For the existence of a real root of the equation \eq{vz}
above, we must have $D>0$.

\begin{figure}
\begin{center}
\includegraphics[width=0.7\textwidth]{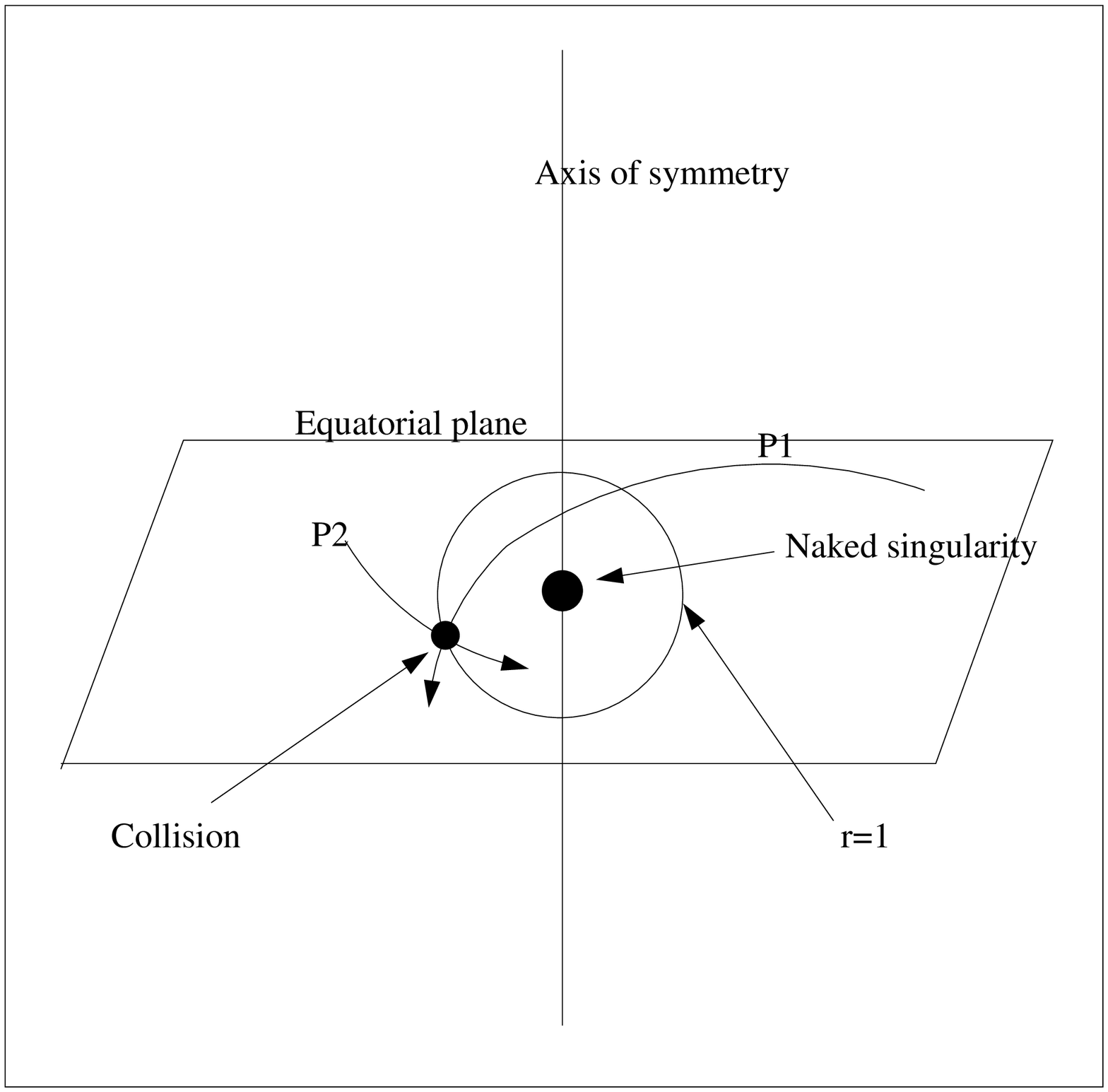}
\caption{\label{fg1}
Schematic diagram of a Kerr spacetime with naked singularity.
One of the particles which is initially ingoing turns
back due to the angular momentum barrier, and it then collides
with another ingoing particle near $r=1$.
Both the particles
follow the geodesic motion in the equatorial plane.
The center of mass energy of collision is arbitrarily
large in the limit $a\rightarrow 1^{+}$
}
\end{center}
\end{figure}

It can be easily seen that
for extremely small values of the angular momentum
$\mid L\mid \rightarrow 0$, $D \rightarrow -\infty$,
the equation \eq{rrefl} does not admit any real
roots. Thus in such a case the ingoing particle
never turns back and it continues its motion
inwards to hit the singularity.

On the other hand, for the
very large values of the angular momenta, as
$\mid L \mid \rightarrow \infty$, we have $D \rightarrow 1$.
Therefore in that case the ingoing particle gets reflected at
an extremely large value of the radial
coordinate $r_{refl} \approx \frac{L^2}{2}$.

In fact, there exists an intermediate critical
value of the angular momentum $L$ for the particle,
which is given by a solution of the equation $D=0$.
This has the property that if the angular momentum is
larger than this critical value, the initially infalling
particle eventually turns back, and if the angular
momentum is smaller than this value, then the particle
will fall inwards and would eventually hit
the singularity.

The angular momentum of the particle can be
oriented either parallel to the spin of a naked singularity
or it could be antiparallel. We first assume that it is
parallel so that $L>0$.

The equation $D=0$ is to be solved in order to
obtain a minimum critical value of the angular momentum
$L_{crit}$ for the particle to turn back. We can write this
equation as,
\begin{equation}
L^2=2|L-a|=\pm 2(a-L)
\label{lll}
\end{equation}
The positive or negative signs in the equation
above stand for the cases where the critical angular
momentum is smaller or larger than the Kerr spin
parameter $a$ respectively.

We note that in the Kerr blackhole case, if we
write the equation \eq{lll} above with a positive sign
and obtain the critical angular momentum from
the same as a solution, then if the ingoing particle were
to turn back, the turning point happens to be necessarily
inside the horizon. Such a scenario is clearly not allowed.
This then implies that in the blackhole case, for
the particle to turn back, the allowed values for the
critical angular momentum must be larger than the Kerr spin
parameter. In that case we have to solve the above equation
with a minus sign.  Then the angular momentum that solves
\eq{lll} yields a legitimate turning point which is outside
the event horizon. Thus in the blackhole case, the
solution to \eq{lll} with a minus sign is the critical
angular momentum for the particle to turn back.

The situation is quite different for a naked singularity,
which corresponds to $a>1$ values, essentially due to the
absence of an event horizon. The legitimate turning point
for an ingoing particle in this case could be in principle
all the way upto the singularity, which is located at
$r=0,\theta=\frac{\pi}{2}$. This is unlike the blackhole
case, where the turning point must be strictly located
outside the horizon as we noted above. Since we are looking
for the smallest value of the angular momentum for which
an initially ingoing particle turns back, we solve \eq{lll}
with a positive sign. It turns out, as we show below, that
the turning point, with the critical angular momentum
obtained by solving \eq{lll} with the positive sign, is
at a positive value of the radial coordinate. Thus, the
allowed critical value of the angular momentum for the particle
to turn back happens to be smaller than the Kerr spin
parameter in this case.

\begin{figure}
\begin{center}
\includegraphics[width=0.7\textwidth]{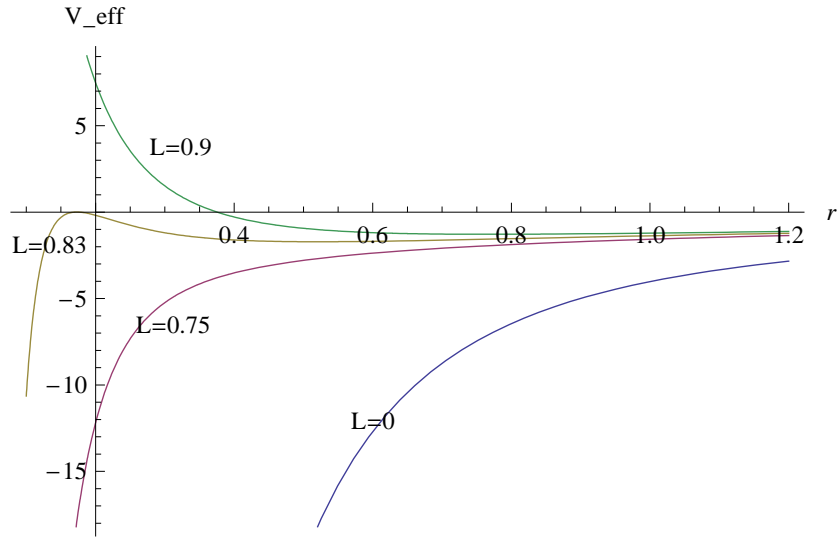}
\caption{\label{fg2}
The effective potential is plotted for a particle
with $E=1$. The Kerr parameter is assumed to be $a=1.005$.
The effective potential is a monotonically decreasing function
for $L=0$, indicating that the particle will travel radially inwards with
ever-increasing radial velocity component and hit the singularity.
The behavior is similar for the particle with subcritical
angular momentum $L=0.75$. The effective potential with the critical
angular momentum $L=0.83$ barely manages to touch the zero
at $r=0.18$. This is the minimum value of the angular momentum
of the particle for which it turns back. The effective potential
admits a zero at $r=0.37$, where the particle turns back, for
a supercritical value of the angular momentum $L=0.9$.
}
\end{center}
\end{figure}

We solve \eq{lll} to obtain
\begin{equation}
L_{crit}=2\left(-1+\sqrt{1+a}\right)
\end{equation}
The turning point for the ingoing particle with
the critical angular momentum $L_{crit}$, will be
\begin{equation}
r_{refl,crit}=\frac{L_{crit}^2}{4}= \left(\sqrt{1+a}-1\right)^2
\label{refl}
\end{equation}
For Kerr naked singularities, since $a>1$, we get a
turning point at a location away from the naked singularity
at $r=0,\theta=\frac{\pi}{2}.$

The fact that the solution to \eq{lll}, with a
plus sign indeed yields a critical angular momentum, which
happens to be smaller than the Kerr spin, is
explicitly demonstrated by Fig2.
The behavior of the effective potential for the particles
with angular momenta smaller, larger and equal
to this critical value are also plotted in Fig2. The effective
potential for the subcritical angular momentum does not admit
a zero, indicating the absence of any turning point. The
effective potential for the critical value of the angular
momentum barely manages to take a zero value. It in fact admits
a maximum. The effective potential for the particle with
supercritical angular momentum cuts the horizontal axis,
from where the ingoing particle can turn back.

We clarify that the turning point $r_{refl,crit}$ for
a particle with the critical angular momentum has the
following sense. The ingoing particle will asymptotically
approach $r=r_{refl,crit}$ as the proper time tends to infinity,
since both the effective potential and its derivative vanish
as it can be seen clearly in Fig2. However, for any value of
the angular momentum larger than this critical value, the
particle turns back. This is because the effective potential
is zero but its derivative takes a nonzero value at
the turning point.

It follows that the angular momentum of the particle
should be strictly larger than the critical value
if it is to turn back, and we have,
\begin{equation}
L > 2\left(-1+\sqrt{1+a}\right)
 \label{Lp1}
\end{equation}
We note that the angular momentum of the particle
to turn back can either be smaller or larger than the
Kerr spin parameter $a$ as long as it satisfies the
condition given above. It is only the critical value of
the angular momentum of the particle which is smaller
than the Kerr spin parameter.

Since we want collisions to take place at $r=1$,
one of the colliding particles must get reflected back
from a radial coordinate $r<1$.
Thus we further impose a condition that
\begin{equation}
r_{refl} < 1
\label{rref}
\end{equation}
The upper limit on the angular momentum of the
particle obtained from the equation above is given by
 \begin{equation}
L< \left(2a-\sqrt{2a^2-2}\right)
\label{Lp2}
\end{equation}
The effective potential for the angular momenta
smaller, larger and equal to the above value are
plotted in Fig3. As is evident from the figure, only the
effective potential of the particles with angular momenta
satisfying the conditions above admits a zero for $r<1$.

\begin{figure}
\begin{center}
\includegraphics[width=0.7\textwidth]{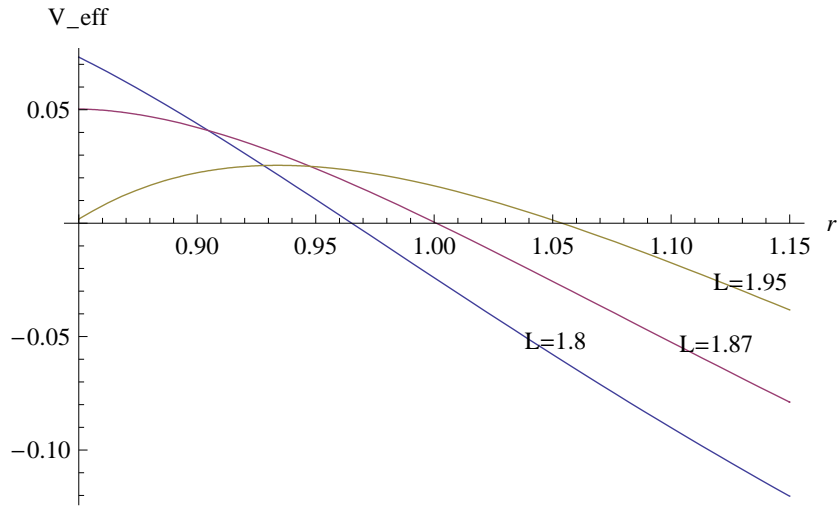}
\caption{\label{fg3}
The effective potential is plotted for a particle
with $E=1$. The Kerr parameter is assumed to be $a=1.005$.
It is clear that the angular momentum for the particle
to turn back at a radial value $r<1$ must be lesser
than the upper bound $L=1.87$, which is the critical value
for which the particle turns back from $r=1$.
For example, for $L=1.8$ the particle turns back from
$r=0.96 <1$, whereas for $L=1.95$, the particle
turns back from $r=1.05$ which is above the
$r=1$ surface.
}
\end{center}
\end{figure}

Combining together the conditions \eq{Lp1} and \eq{Lp2},
we obtain the interval of the allowed angular momenta
values of one of the particles which is initially ingoing and
later turns back, as below,
\begin{equation}
 2\left(-1+\sqrt{1+a}\right)< L < \left(2a-\sqrt{2a^2-2}\right)
\label{Lr}
\end{equation}

The only condition that must be imposed on the
second particle is \eq{Lp2}, so that it does not get
reflected back at $r>1$ and actually reaches $r=1$
as an ingoing particle.

Therefore, the particle dropped in from infinity,
which moves along the equatorial plane with angular
momentum in the range given by \eq{Lr}, crosses $r=1$
as an ingoing particle, and it is then reflected
back at the radial coordinate $r<1$. It then again
reaches $r=1$ as an outgoing particle, where it
interacts with another ingoing particle dropped
from infinity at rest.

The proper time required for this process to
occur happens to be finite, since both the conditions
mentioned in \eq{Veffz} are not satisfied simultaneously
anywhere along the geodesic.

When the angular momentum of particles is
oriented antiparallel to the spin of the naked singularity
with $L<0$, it can be shown that the simultaneous solution
to \eq{vz} and \eq{rref} does not exist. Thus such
particles are not useful towards the purpose
of high energy collisions.

The center of mass energy of collision between
these two particles is computed using \eq{Ecm}. It requires
the calculation of the inner product of velocities of
the two particles. The velocities of two particles are
given by \eq{Ut},\eq{Up},\eq{Ur} with $E=1$, and
which have appropriate angular momenta as discussed above.
The expression for the center of mass energy of collision
contains terms that are proportional to $\frac{1}{\Delta}$,
those which are independent of $\Delta$, and others with
positive powers of $\Delta$. Since the Kerr spin parameter
is very close to unity, it follows from \eq{del} that
the terms proportional to $\frac{1}{\Delta}\approx
\frac{1}{2\epsilon}$ would make a dominant contribution,
and the other terms can be neglected being insignificantly
small as compared to it. The center of mass energy of collision
to the leading order is then given by,
\begin{equation}
\lim_{\epsilon \rightarrow 0} E_{c.m.}^2=2 m^2
\frac{\tilde{T_{1}}\tilde{T_{2}}}{\epsilon} \rightarrow \infty
\end{equation}
where $\tilde{T_{1}}=T_1 (r=1,L_1,E=1),\tilde{T_{2}}=T_2
(r=1,L_2,E=1)$, and the functions $T_{1},T_{2}$ are defined
below \eq{Up} in Sec 2.
We have, $\tilde{T_{1}},\tilde{T_{2}}\approx O(1)$.
Thus we clearly see that the center of mass energy
of collision between two particles is arbitrarily
large in the limit where the deviation
of a Kerr naked singularity from extremality is small.

\section{Discussion and open issues}

We first note that the consideration of Kerr naked
singular geometries is well-motivated by recent theoretical
developments in string theory, which suggest by means
of specifically worked out examples, that the timelike naked
singularities would be naturally resolved. Possible
pathological features associated with them like causality
violation would be naturally avoided by the high energy
modifications to classical general relativity, and
predictability would be restored
\cite{Gimon}.
In such a case, the cosmic censorship conjecture
\cite{Penrose}
which forbids the existence of the naked singular
solutions in nature becomes obsolete. Quantum gravity
resolved classical naked singular solutions are then
rendered legal and can be used to do calculation as far
as one stays sufficiently away from the
ultra-high curvature regime.

The results obtained in this paper basically illustrate
the mathematical structure of the Kerr geometry.
For the situation we described here to be astrophysically
relevant, various other issues need to be addressed.
We qualitatively discuss some of these points in this section.
Rigorous analysis of these questions and all the
open issues is beyond the scope of this paper and will
be presented elsewhere in future.

The formation of spherically symmetric naked singularities
in gravitational collapse has been studied extensively.
There are many such collapse models where the violation of
cosmic censorship conjecture occurs
\cite{Joshi}.
On the contrary, the formation of either rotating
blackholes or naked singularities from gravitational collapse
is not yet very well understood.
The formation of both rotating naked singularities
and blackholes in gravitational collapse of $2+1$ dimensional
shell collapse has been demonstrated recently
\cite{Mann}.
It was also shown that the accretion onto compact objects
can spin these up to super-spinning configuration.
However, this also requires the compact object to have an
analogous quadruple moment apart from mass and angular momentum
\cite{Bambi0}.
It has also been suggested that the near extremal
Kerr blackhole can be turned into a naked singularity by
throwing in test particles
\cite{Jacobson},
although it is a matter of debate and investigation
whether the results would survive after the self-force and
backreaction have been taken into account.

The Kerr naked singular solution is not the unique vacuum,
asymptotically flat, axially symmetric solution to the Einstein
equations, the most general solution being the Tomimatsu-Sato
geometries
\cite{Sato}.
We analyzed here the simplest subcase, namely the
Kerr naked singular solution. We are currently investigating
whether the high energy collisions can also take place in
the spacetimes with analogous higher multipole moments.
Yet another variant of a Kerr naked singular solution is
the non-vacuum, axially symmetric, asymptotically flat solution
with the massless scalar field as a matter source
\cite{Krori}.
We have verified that the high energy collisions do take
place in this geometry admitting rotating naked singularities.
Thus it might be reasonable to expect that the results we
presented here would carry over to geometries
other than the Kerr naked singularity.

We carried out the analysis here under the assumption
that the colliding particles are test particles and followed
a geodesic motion on the background Kerr geometry. However,
in the full calculation, the backreaction and gravitational
radiation emitted by infalling particles must be taken into account.
In the blackhole case, one of the colliding particle follows
a whirl orbit. It is an orbit which asymptotes to the horizon and
the particle circles around the horizon many times. Such a particle
has a fine-tuned value of the angular momentum. This particle
emits a large amount of gravitational radiation and its orbit
suffers a severe deviation, thereby reducing the center of mass
energy by a large amount
\cite{Berti}.
In the process we described here, fine-tuning of the
angular momentum is avoided. Thus the gravitational radiation
emitted is significantly reduced, since the particle trajectory
is not a whirl orbit. Although it is expected that the particle
would be deviated from a geodetic motion, since all we need is an
ingoing and outgoing particle colliding around $r=1$, it is
expected that there would be high energy collisions. Secondly,
if one assumes that the particles would be accreted in the form of
quasi-spherical shell then the gravitational radiation per particle
would be much lesser than the gravitational radiation emitted by
a single infalling particle. The issue of backreaction is as such
difficult to deal with here in the absence of spherical symmetry.
The full general relativistic calculation, taking into account
the backreaction, was carried out in
\cite{Nakao}
and it was shown that the center of mass energy of collision
between the shells around the extremal Reissner-Nordstr\"{o}m blackhole
turns out to be finite, as opposed to the unbound center of mass
energy, when the test shell approximation is used. Contrary to the
above result, we have shown recently that in the case of naked singular
Reissner-Nordstr\"{o}m spacetime the center of mass energy of collision
turns out to be unbound even when the exact calculation is
carried out taking into account the backreaction
\cite{Patil}.

We also note that the particles released from rest at
infinity are highly blueshifted as they reach $r=1$, where they
participate in the high energy collisions with extremely large
center of mass energy. The high energy particles produced in the
collisions will be highly redshifted when they reach out infinity.
But there is an overall compensation of blueshift and redshift
so that the particles reaching infinity carry energy that is
comparable to the mass of the infalling particles. This can also
be argued in the following way. The conservation of energy-momentum
in the collision implies that the total conserved energy of the
colliding particles would be same as the total conserved energy
of the collision products. Thus the conserved energy of the particles
produced in the collision escaping to infinity would be of the
order of the conserved energy of the colliding particles. The
conserved energy is the energy of the particle measured at infinity.
Thus the particle produced in the collision escaping to infinity
will carry the energy that is comparable to the mass of the colliding
particle if it is released from rest at infinity. If the mass of
the particle is much lesser than the mass of the colliding particle
then the particle produced in the collision will reach infinity
with large kinetic energy.

In general the stability of a given spacetime, including
either Kerr blackholes or naked singularities, is an open issue
and is still under investigation. It has been demonstrated
recently that the Kerr naked singular solution admits an
instability
\cite{stabilityNS}.
We note that Kerr blackholes also could admit
certain instabilities
\cite{Jacobson},\cite{stabilityBH}.
These are different kinds of instabilities associated
with the Kerr solution in standard general relativity
and also for the $F(R)$ gravity theories. Recently yet
another instability is under much discussion, which is that
associated with the near extremal blackholes, in the case
when the blackhole absorbs charged or rotating particles,
possibly turning it into a naked singularity as we
discussed above. These instabilities could play a vital role
in particle acceleration mechanism associated with the
near extremal blackholes. Despite that the extremal blackholes
have been extensively studied from the perspective of
particle acceleration mechanism in past couple of years.

We note that more careful further analysis is needed
to investigate whether or not, and under what circumstances
the process we described here will be stable when the
back-reaction effects and the gravitational radiation emitted
by particles is taken into account. Since this issue is
complex one to deal with in Kerr geometry, we have done an
exact analogous calculation taking into account the
full backreaction, in the Reissner-Nordstr\"{o}m naked singular geometry.
It turns out in that case that the Reissner-Nordstr\"{o}m geometry is stable
and the center of mass energy of collisions can be arbitrarily
large. So it might be reasonable to expect that similar
results might hold good in Kerr geometry as well, possibly
under certain restrictive conditions.

\section{Concluding remarks}

To summarize, we described a process of high energy collision
of particles in the vicinity of near extremal Kerr naked singularities
which is generic and requires a finite proper time, unlike in
the Kerr blackhole case. The genericity is with respect to the finite
range and interval of values of angular momenta that is available
to particles that participate in the high energy collisions.
This is unlike the blackhole case where extreme finetuning of
angular momentum is required.

This is in itself an intriguing and interesting result
that it is possible to have collisions with large center of mass
energies around the Kerr naked singularities. However, for this
phenomenon to be physically relevant, it is
important to study and understand issues like the possible processes
leading to the formation of Kerr naked singularities and the
deviation of the colliding particles from geodesic motion due
to the gravitational radiation and also the backreaction
effects.

\end{document}